\documentclass[graybox, nosecnum]{svmult}

\usepackage{mathptmx}       
\usepackage{helvet}         
\usepackage{courier}        
\usepackage{type1cm}        
\usepackage{makeidx}         
\usepackage{graphicx}        
\usepackage{multicol}        
\usepackage[bottom]{footmisc}
\usepackage{hyperref}        
\usepackage{soul}            

\makeindex             

\begin{document}
\title*{Beta decay of halo nuclei}

\author{Karsten Riisager}

\institute{Karsten Riisager \at Department of Physics and Astronomy, Aarhus University, DK-8000 Aarhus, \email{kvr@phys.au.dk}}
\maketitle
\abstract{Beta decay is a well-established and efficient probe of nuclear structure and provides important information also for halo nuclei. Detailed decay studies of near-dripline nuclei are challenging, but are now feasible due to the technical progress during the last decades in isotope production as well as detection capabilities. The halo structure can change the dynamics of the decay process, most noticably when decays proceed directly to continuum states as seems to be the case in beta-delayed deuteron emission, a process only observed until now for the halo nuclei $^6$He and $^{11}$Li. The pronounced clustering in halo states also leaves an imprint on the decay patterns, although the isospin symmetry remains important; in several cases the isobaric analogue states of halo states have also been established. Beta-delayed particle emission channels can be expected to dominate the decays of heavier halo nuclei, if not in terms of branching ratio than certainly in terms of beta strength. A remaining challenge is to find experimental as well as theoretical tools to study the multi-neutron final states that appear for most neutron dripline nuclei.}

\section{Introduction}
A general overview of halo structures in nuclei, as well as historical accounts of how this research field has evolved, can be found e.g. through the following review papers: \cite{Fre12, Han95, Jen04, Rii94, Rii13, Tan96, Tan13}.  As a very brief summary many features of what we now consider halo physics were known early on for the deuteron that can be thought of as the lightest nuclear halo state. However, it was only in the mid 80's through the papers \cite{Tan85} and \cite{Han87} that it was realized that other nuclei with low nucleon separation energy in their ground states could display similar - and in some cases larger - effects and the halo structure obtained its name. The low separation energies and the large spatial size give a particular dynamics, often with striking single-particle features. There is by now strong evidence for this from nuclear reactions at several energy scales, as well as confirmation from other experiments.

The deuteron does not beta decay, but all other ground state halos will do so. The aim of the current chapter is to present to what extent the dynamics of beta decay is affected by a halo structure and how beta decays can provide information on halos. This line of investigation was triggered early on by the suggestion of the late Jens Lindhard \cite{Han87a} that a two-neutron halo e.g. in $^{11}$Li may beta decay into a deuteron with the core as a more-or-less non-interfering spectator, a process not observed at the time. Now, many experiments and theoretical papers later, beta decay is established as a valuable probe of nuclear halo structure.

The exposition of this is organized as follows: first, two brief subsections summarize the most important features of halo structures and beta decays. The following section outlines the main ways halo structures affect beta decay and includes specific discussions of beta-delayed particle emisson branches and of the role of isospin. Two brief sections give an overview of which theoretical approaches and experimental procedures have been found valuable so far. The two following sections give an overview of the current knowledge of beta-decays of neutron halo nuclei, proton halo nuclei and hypernuclear halos. Finally, the main findings are summarized.

\subsection{Halo nuclei}
The first striking feature of halo nuclei was their large spatial extension. This occurs when one or two nucleons are bound with a low separation energy and therefore are able to tunnel a substantial distance out from the binding potential. However, formation of a halo state is not automatic when separation energies are low. A key requirement is that the structure of the state should resemble the one of the threshold it lies at, i.e.\ a significant fraction of its wavefunction should have the form of a core and one or two nucleons. Whether such cluster structures occur preferentially at thresholds is still being investigated; it happens frequently in light nuclei, but not always.

The tunneling out of the nuclear potential will be hindered by angular momentum barriers or Coulomb barriers. The former limits prominent halo formation to nucleons in s- and p-waves, the latter prevents sizable proton halo effects in heavier nuclei. (The situation is very similar for slightly unbound proton states where a strong Thomas-Ehrman shift is only seen for large spectroscopic factors and low orbital angular momentum, see e.g.\ \cite{Bar91}.)
For an $N$-body system a barrier in a radial equation very similar to a angular momentum barrier will appear with effective angular momentum $(3N-6)/2$. This in principle limits halo formation to $N=2,3$ (one and two nucleon halos), although correlations between particles can ameliorate the situation.

For an s-wave neutron with separation energy $S$ the radial wavefunction outside the potential will be of the Yukawa type
\begin{equation}  \label{eq:Yukawa}
    \frac{1}{r} \exp(-\kappa r) \; , \; \kappa = \sqrt{2 m S / \hbar^2}
\end{equation}
where $m$ is the reduced mass. For small binding the root-mean-square radius will diverge as $S^{-1/2}$, but it is noteworthy that the overlap with the core remains sizeable for all known nuclear halo states.

The currently established halo nuclei have mass numbers below 40. There are different suggestions for how the halo concept may be extended to heavier nuclei, but it will take time before this can be tested experimentally.

\subsection{Beta decay}
There is a large literature on beta decay in textbooks as well as in specialized reviews. The brief overview here serves mainly to remind on the main features and to present the notation used in later sections. A more detailed general treatment can be found in books such as \cite{Beh82, Hol89} and more specific accounts on beta decay in exotic nuclei are given in the reviews \cite{Bla08,  Jon01, Pfu12}.

For a transition between nuclear states with spin-parity $\vec{J_i^{\pi_i}}$ and $\vec{J_f^{\pi_f}}$ that involves leptons with total spin $\vec{S}$ and orbital angular momentum $\vec{L}$, conservation of parity and angular momentum gives the relations $\pi_i = \pi_f (-1)^L$ and $\vec{J_i} = \vec{J_f} + \vec{S} + \vec{L}$. We shall mainly be interested in allowed transitions ($L=0$). The selection rules for Fermi ($S=0$) / Gamow-Teller ($S=1$) transitions then are: no parity change and $\Delta J$ equal to 0 / 0 or 1, respectively. The nuclear operators for allowed Fermi and Gamow-Teller $\beta^{\pm}$ transitions are
\begin{equation}  \label{eq:operators}
   F_{\pm} = \sum_i t^i_{\pm} = T_{\pm} \,,\;  GT_{\pm} = \sum_i \sigma^i t^i_{\pm}
\end{equation}
where the sums are over all nucleons, $t$ is the nucleon isospin operator (with the convention that neutrons have isospin projection $+1/2$ and protons $-1/2$), $T$ the total isospin of the nucleus, and $\sigma$ a spin Pauli matrix (there are three matrices and therefore three components of the Gamow-Teller operator). The squares of the matrix elements give the beta strength $B(F)$ or $B(GT)$. The sum rules for the total beta-strength for transitions from a given nuclear state are
\begin{equation}
  \sum_j B(F)^-_j - \sum_k B(F)^+_k = N-Z \;, \;
   \sum_j B(GT)^-_j - \sum_k B(GT)^+_k = 3(N-Z) \;.
\end{equation}
Essentially all Fermi strength will go to the Isobaric Analogue State (IAS) that will be discussed further below. The Gamow-Teller strength is concentrated in the Gamow-Teller Giant Resonance (GTGR) that typically is situated above the IAS, but with wide tails, so that most observed beta transitions will be Gamow-Teller transitions.

The basic relation connecting the strength of a given $\beta^{\pm}$ transition with the $ft$-value, i.e.\ the relation between the theoretical and experimental quantities, is
\begin{equation}   \label{eq:ft-relation}
  ft = \frac{\mathcal{T}}{B(F) + \left(\frac{g_A}{g_V}\right)^2B(GT)} \;,\;\;
  \mathcal{T} = \ln 2 \frac{2 \pi^3\hbar (\hbar c)^6}{g_V^2 (m_e c^2)^5}  = 6144(4) \mathrm{s}
\end{equation}
where $g_V$ and $g_A$ are the vector and axial-vector weak coupling constants with the experimental ratio $g_A/g_V = -1.2754(13)$ \cite{PDG20} and the experimental value for $\mathcal{T}$ is taken from superallowed Fermi decays \cite{Har20}.
There are different definitions of beta strength in the literature, in some the Gamow-Teller strength includes the factor $g_A/g_V$.
The corresponding relation for electron capture (EC) involves a sum over capture from different atomic orbitals where each term has a phase space factor proportional to the square of the energy $E_{\nu}$ of the outgoing neutrino.

The $t$ on the left-hand side of eq. (\ref{eq:ft-relation}) is the partial halflife for the transition in question, and $f$ is the phase space factor that is dimensionless and expressed in terms of $p' = p_0/(m_ec)$ and $E' = E_0/(m_ec^2)$ where $E_0$ is the maximum (relativistic) electron energy for the transition and $p_0$ the corresponding momentum.  For $\beta^+$ transitions $E_0 = Q_{EC}-m_ec^2$, and for $\beta^-$ transitions $E_0 = Q_{\beta}+m_ec^2$. Neglecting Coulomb effects on the outgoing leptons one has
\begin{equation}
  f_{Z=0} = \frac{1}{60} \left( 2E'^4 - 9E'^2 -8 \right) p' + \frac{1}{4}E' \ln (E'+p')
             \approx \frac{1}{30} \left( \frac{E_0}{m_ec^2} \right)^5
\end{equation}
where the last expression singles out the leading term at large transition energy. The different powers of the energy makes $\beta^+$ decays dominate EC decays as the transition energy increases. A very complete account of the allowed $\beta$ spectrum shape is given in \cite{Hay18}.

\section{Aspects of halo beta decay}
The beta strength depends on the structure of both initial and final state and will thereby reflect the structure of both states. We shall here be concerned with features of the beta decay that are clearly related to the halo structure and can for presentational purposes divide this into four aspects.

\begin{figure}[thb]
\centering
    \includegraphics[width=15.cm,clip]{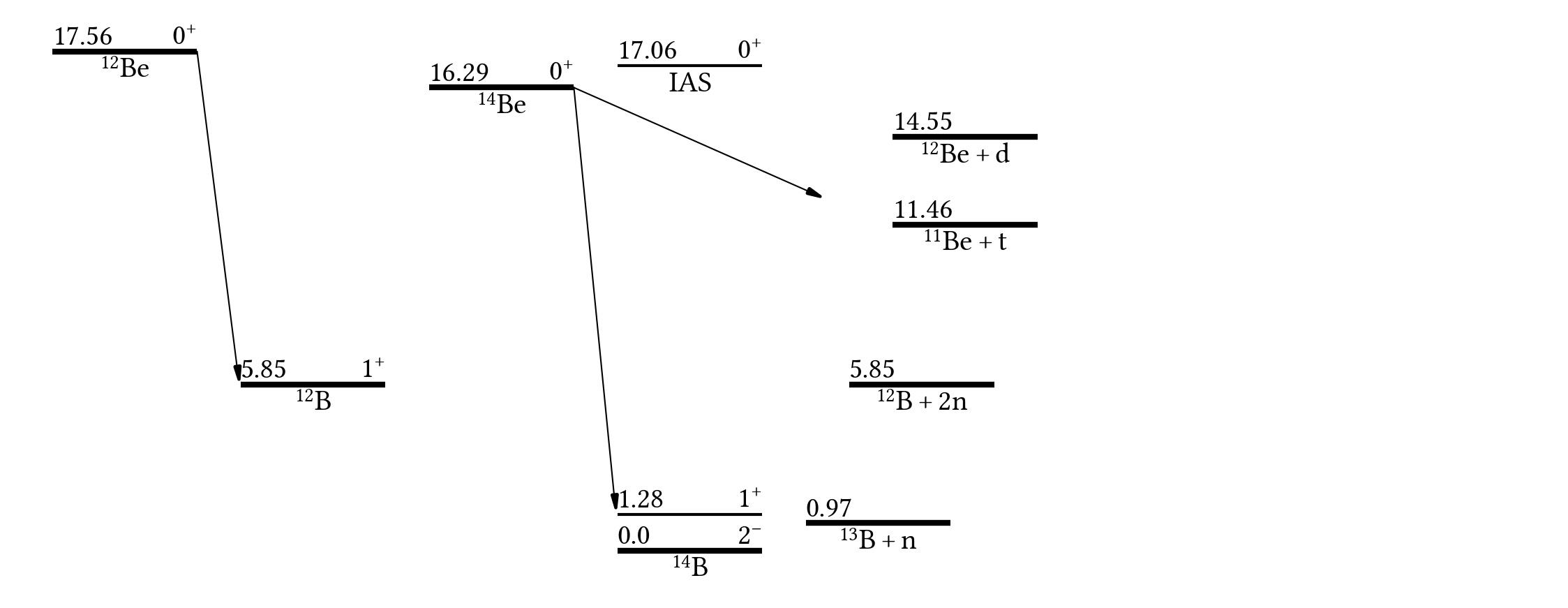} 
    \caption{The beta decay of $^{14}$Be leads mainly to a $1^+$ excited state in $^{14}$B, very similar to the decay of $^{12}$Be to the ground state of $^{12}$B. Some of the thresholds for delayed particle emission are indicated.}
\label{fig:Be1214scheme} 
\end{figure}

As a \emph{first aspect}, halo nuclei above mass number six have large Q-values for beta decay, from around 10 MeV to more than 20 MeV, and since the daughter nuclei also tend to have low separation energies beta-delayed particle emission will be prominent. An example is given in figure \ref{fig:Be1214scheme} that shows that beta-delayed emission of one neutron, two neutrons, tritons and deuterons (and also three neutrons and alpha particles) in principle is allowed for $^{14}$Be. Beta-delayed particle emission will be discussed in a subsection below, more information can be found e.g.\ in \cite{Pfu12}.

The \emph{second aspect} is the effect of the cluster structure of halo states that allows them to be described in a first approximation as a core and a surrounding halo. When the total wavefunction factorizes into a core part and a halo/``few-body'' part describing the halo particles relative to the core, the beta decay process formally simplifies. The beta-decay operators in equation (\ref{eq:operators}) are linear and the final state in the beta decay therefore will have two components \cite{Nil00, Jon04}: a core decay component where the halo is a spectator, and a halo decay component with the inert core remaining. The decay of $^{14}$Be shown in figure \ref{fig:Be1214scheme} would then be interpreted as being dominated by a core decay branch. Although this way of thinking about the decay neglects isospin and thus cannot be correct (the isospin aspects are discussed in more detail below), it may give a natural way of interpreting patterns in the decay.

Clustering in general in nuclear physics is still a very active research field, see \cite{Fre18} for a recent overview. It is often noticed in states close to the threshold for break-up into the cluster components, where it will have a large influence on the dynamics of reactions as well as decays. The loosely bound neutron in the deuteron is often used as a replacement of a pure neutron target, and the decay of halo neutrons may in a similar manner be thought of as closely following the decay of free neutrons (as used e.g.\ in connection with the recent suggestion of a dark decay channel for neutrons \cite{For18}).

The \emph{third aspect} is connected to the large spatial extension of halo states. Taken by itself this will reduce the spatial overlap to ``normal'' final states. Due to the Yukawa  type wavefunction, eq. (\ref{eq:Yukawa}), that is rapidly decreasing at large radii, but of wide total extent, the decrease in overlap does not scale with the rms-radius and is often a moderate effect. The effect on the transition rate to a state in the daughter nucleus may thus be hard to identify unless good theoretical predictions are available. However, the Yukawa wavefunction tail will have enhanced overlap with continuum states of the final system, so decays directly to the continuum may occur. It is not always straightforward to experimentally distinguish decays through resonances from decays directly to the continuum; this question of decay mechanism is treated in more detail in the subsection on beta-delayed particle emission. As shown there, certain halo beta decay branches are indeed most naturally explained as proceeding directly to the continuum.

The \emph{fourth aspect} is that beta decay can give valuable information on the structural composition of the halo state. This is in line with the standard use of beta decay as a probe of nuclear structure, but of course adds important information when combined with that derived from nuclear reactions or other processes involving halo nuclei. An early example is the determination of the parity of the ground state of $^{11}$Be as being positive \cite{Alb64}, a more recent example is the calculation \cite{Suz94, Suz97} of the decay rate of $^{11}$Li into the first excited $1/2^-$ state of $^{11}$Be that is sensitive to the configuration of the two halo neutrons in $^{11}$Li, in s-orbits, p-orbits etc, since the $1/2^-$ state will favour the p-orbit component. In this specific case the effect is sufficiently large to be also visible in calculations \cite{Bar77} of the total halflife of $^{11}$Li.

The following smaller points should also be kept in mind:
\begin{itemize}
\item Even though the above discussion implicitly assumes the halo state to be the initial state, halo structures in excited states in nuclei can of course also be probed if fed in beta decay.
\item In light very neutron-rich nuclei a significant part of the GTGR is lowered and enters the accessible Q-window so that strong Gamow-Teller transitions often are seen in the decays \cite{Bor91}. This appears to be a general phenomenon and not coupled to halo formation \cite{Sag93}. However, it does allow more of the beta strength to be experimentally accessed in nuclei close to the neutron dripline.
\item So far mainly allowed transitions have been discussed. The operators for first (and higher order) forbidden decays can enhance the sensitivity to the outer parts of the wavefunctions compared to allowed transitions. In this respect they resemble radiative E1 (or higher order) capture that will be sensitive both to proton and neutron halos \cite{Ots94, Rii92}. The inverse process, photodisintegration, is of course an established probe in itself of ground state halo states.
\end{itemize}

\subsection{Beta-delayed particle emission}
All halo nuclei (except the deuteron) will have $Q$-values for beta decay that are higher than particle separation energies in the daughter, and beta-delayed particle emission is indeed a prominent feature for all known halo nuclei above $A=6$. Some of the delayed emission processes are closely related to the halo structure as is seen by rewriting \cite{Jon01} the $Q$-values for the processes for $\beta^-$ decays of neutron halos:
\begin{equation}
  Q_{\beta d} = 3007 \,\mathrm{keV} -S_{2n} \;,\; Q_{\beta p} = 782 \,\mathrm{keV} -S_n,
\end{equation}
and correspondingly for EC decays of proton halos:
\begin{equation}
  Q_{EC d} = 1442 \,\mathrm{keV} -S_{2p},
\end{equation}
where all separation energies are the ones for the initial (halo) state.
These relations suggest an interpretation of the decays as two halo-neutrons decaying into a deuteron etc. The alternative core decay can also become of interest, as in the case of the beta-delayed proton decay of $^8$B where the halo proton is a spectator that is unbound after the decay.

\begin{figure}[thb]
\centering
    \includegraphics[width=10.cm,clip]{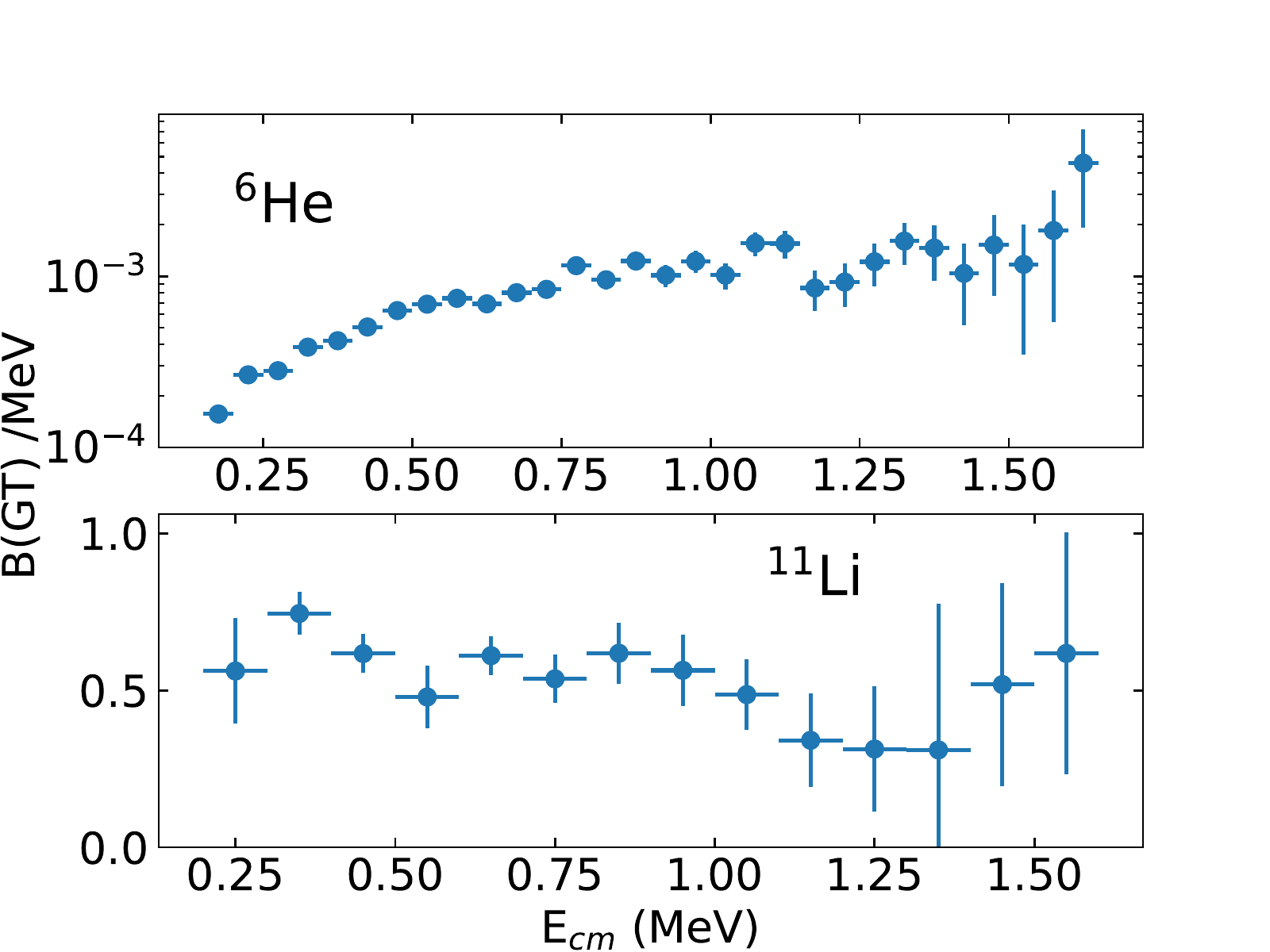} 
    \caption{The experimental $B(GT)$ strength for the beta-delayed deuteron decays of $^{6}$He and $^{11}$Li. The strength has been converted through eq.\ (\protect\ref{eq:ft-relation}) from the data in refs \protect\cite{Pfu15,Raa08}. Note that a logarithmic scale is employed for the strength of $^6$He and a linear scale for $^{11}$Li.}
\label{fig:BGT_betad} 
\end{figure}

The beta-delayed deuteron decay is the decay mode where the best indications for halo effects have been obtained. Two cases have been established experimentally, $^6$He and $^{11}$Li, and the beta strength distribution to the deuteron channel for these two cases is shown in figure \ref{fig:BGT_betad}. The first to be measured was $^6$He, it turns out (as discussed in detail below) that the main effect of the spatial extension in this decay is to suppress the intensity of the deuteron channel by two orders of magnitude so that the observed integral strength is $1.6 \cdot 10^{-3}$, only. 
In contrast, the total observed strength in the case of $^{11}$Li is 0.73 and the shape of the beta-delayed deuteron strength is almost featureless with no clear indication for resonance structure and therefore more consistent with decays directly into the continuum than with decays proceeding through a high-lying resonance in $^{11}$Be.

The remainder of this subsection will treat the question of the decay mechanism in more detail.
Two different approaches are used to describe beta-delayed particle emission: as happening directly to continuum states or as proceeding through resonances in the beta daughter nucleus. Most treatments of beta decay implicitly assume discrete final states and therefore conform with the latter approach, but calculations in potential models, e.g. few-body continuum calculations, explicitly deal with final state continuum states. As discussed below, most theoretical work on $^6$He($\beta$d) has taken place assuming the decay proceeds directly to the continuum. Several groups have used this approach that has been explored in particular by the ULB group \cite{Bay94,Bay06,Des92,Tur18} that also has explored other processes, among them beta-delayed proton emission from one-neutron halo nuclei where the best case is found to be $^{11}$Be \cite{Bay11}.

The traditional approach assuming beta decays into discrete final states is both natural and clearly applicable to beta-delayed particle emission when the final state resonances are narrow and non-overlapping. To deal with situations with overlapping and/or broad resonances more sophisticated frameworks must be employed, the one mostly used is R-matrix theory as adapted to this case by F.C.\ Barker \cite{Bar69,Bar88}. Even in the case of a single isolated resonance the usual Breit-Wigner expression is modified slightly to
\begin{equation}
  \frac{\Gamma(E)/2}{(E_0+\Delta(E) - E)^2+ \Gamma(E)^2/4}  \; , \;\;
  \Gamma(E) = 2 P(E) \gamma^2 \;,
\end{equation}
where $E_0$ is the resonance energy, $\Gamma$ the level width that now is explicitly energy dependent due to the penetrability factor $P$, $\gamma$ is the reduced width amplitude that quantifies how strongly the resonance couples to the continuum and $\Delta$ is the level shift that depends on $\gamma$ as well. This expression describes how broad resonances will become asymmetric and even allows for sub-threshold levels to have a tail above threshold ($P$ being zero below and positive above). Chapter 4 in \cite{Rol88} gives a clear account of these aspects of resonance behaviour as well as the implications in nuclear astrophysics; a good experimental example of the role of sub-threshold resonances is given by the decay of $^{16}$N \cite{Buc93} (it also illustrates the clear role interference will have when resonances overlap). This phenomenon turns out to be crucial for the decay of $^6$He.

In cases where populated resonances interfere, one clearly has to tread carefully when extracting the beta strength of the decay. For complex experimental spectra it can be difficult to obtain fits that are good and unique (in the sense that all relevant levels and decay channels have been identified and their contribution can be separated). More work is probably needed to clarify the situation; the following procedure looks at the moment to be the most reliable one:

The basis is the careful discussion by Berggren and others \cite{Ber68,Ber93} on how resonances and continuum states may be treated. Starting from a pure continuum description one may introduce resonances if they are defined carefully, but the strength attributed to a given resonance will not always be positive, and one will not always be able to distinguish clearly between resonances and the continuum. Particularly useful is the work on sum rules in this situation \cite{Ber73,Rom75}. It indicates \cite{Rii14} that the beta decay sum rules can be maintained most transparently if one stays with the continuum view, i.e. that one defines the beta strength in each energy bin from eq.\ (\ref{eq:ft-relation}) to obtain the overall strength distribution. The beta strength to a given resonance is then only well defined in cases where the resonance is isolated, which may hamper comparison with theoretical strength calculations that assume discrete final states. This should be thought of not as a shortcoming of the Berggren picture, but rather as a reflection of the complexity of the problem.

\subsection{Isospin}
Isospin is an important symmetry in nuclei and as seen from the structure of the operators for allowed transitions, eq. (\ref{eq:operators}), very relevant for beta decay. Isospin is to a good approximation conserved in strong interactions, but is broken by Coulomb interactions. Considerable attention has been given to the IAS of halo states. Let the isospin composition be denoted by $| T, T_3 \rangle$, so that the cluster structure for a neutron-rich halo nucleus gives the combined wavefunction $| T^c, T^c \rangle | T^h, T^h \rangle$ where both core and halo parts have maximum $T_3$ values. The IAS then has isospin $T^{tot} = T^c+T^h$ and the structure
\begin{equation}
  | IAS \rangle = \alpha | T^c, T^c-1 \rangle | T^h, T^h \rangle   +
                          \beta | T^c, T^c \rangle | T^h, T^h-1 \rangle
\end{equation}
with $\alpha$ and $\beta$ being the appropriate Clebsch-Gordan coefficients, see e.g.\ \cite{Suz91} that gave the first predictions for the IAS of $^{11}$Li. The structure follows from applying the operator $T_-$ to the halo state, but can also be seen to arise dynamically \cite{Rob65} due to the isospin-isospin terms in the nuclear potential (e.g.\ from the long-range one-pion exchange). The two-component structure is only avoided if the core has isospin $T^c = 0$.

The IAS of proton halos will be accessible in beta decay, but not that of neutron halos. Here charge-exchange reactions can be employed to find both the IAS and (depending on the kinematic conditions) the GT strength including the GTGR. Two such experiments at RIKEN have already given experimental information on the isobaric analogue states of  $^{11}$Li \cite{Ter97} and $^{14}$Be \cite{Tak01}, a further experiment \cite{Sat11} agrees nicely with the low-energy GT strength from $^{14}$Be. 

Each of the components of the IAS of a halo state will have a cluster structure, but the separation energies will differ so the spatial extensions may differ as well. Nevertheless, a halo structure changes the Coulomb displacement energies compared to states of normal structure \cite{Suz98,Sag15}. Continued operation with the $T_{\pm}$ operators will eventually give the mirror nucleus, and no known halo nucleus has a mirror nucleus that also has a halo structure -- for neutron halos the mirror is unbound, for proton halos the mirror is too bound. In this case, one must be careful not to ascribe all observed mirror asymmetries to the proton halo structure, as the low separation energy also may give differences in orbital occupation numbers in the two mirror nuclei. This will be relevant in the section on decay of proton halos.

A general overview of the isospin structures around halo nuclei can be found in \cite{Izo17}. In addition to the IAS one often discusses the anti-analogue state (AAS) that is orthorgonal to the IAS 
\begin{equation}
  | AAS \rangle = \beta | T^c, T^c-1 \rangle | T^h, T^h \rangle  -
                          \alpha | T^c, T^c \rangle | T^h, T^h-1 \rangle
\end{equation}
and has a total isospin $T^{tot}-1$. Due to the lower isospin value the AAS should not be fed in Fermi decays, whereas Gamow-Teller decays to the AAS can be expected. The very similar structure implies that the Coulomb interaction tends to mix the IAS and AAS slightly. 

The main deviations from isospin symmetry due to the Coulomb interaction are expected to be seen in the low-lying continuum \cite{Rob65}, but it is natural to look now also at the very slightly bound halo nuclei. Calculations for halo nuclei \cite{Suz91,Han93} indicate isospin breaking an order of magnitude higher than in more stable nuclei, but still a very small effect. Note, however, that this is the amount of mixing between the IAS and states of lower isospin. The expected non-perfect radial overlap (due to the different separation energies) of the halo state and the IAS should result in a spreading of the Fermi strength. This is a well-known effect, studied in detail in $0^+ \rightarrow 0^+$ decays. An account within R-matrix theory that explicitly includes breakdown of charge symmetry in the nuclear wave functions is given in \cite{Bar92}.

\section{Theoretical approaches}
A general overview of theoretical models used in the description of halo nuclei can be obtained from the reviews listed in the introduction as well as the specialized review \cite{Sag15}. The question of the decay mechanism in beta-delayed particle emission where continuum final states occur was discussed in the previous section. The current section focusses on how structure effects in beta decay process can be predicted. The different frameworks that have been employed will be briefly listed and contrasted rather than explained in detail.

The interacting shell model \cite{Bro01,Cau05,Ots20} was for many years the standard tool to elucidate the structure of light nuclei and was also a reference for the first attempts of treating beta decays in halo nuclei. There are well-known shortcomings in treating the large-distance behaviour and the coupling to continuum degrees of freedom, specialized developments such as the Gamow shell model succeed in overcoming some of these problems \cite{Mic09}. Where shell models excel is in comparisons between neighbouring nuclei and in tracing structure developments. This is essential for correct interpretation of many beta decay studies.

Mean field and EDF calculations will be very valuable when heavy halo nuclei are reached. Several structure studies have already been undertaken, e.g.\ \cite{Rot09}, and can be extended to include also the beta decay process.

Few-body models have been used extensively to study the structure of the currently established (light) halo nuclei and they describe very efficiently their clustering and spatial configuration as well as the coupling to continuum states. However, many such models require special care in order to simultaneously reproduce the initial and final states in a beta decay. In the discussions below on the individual halo decays, references will be given to some of the many results that have appeared so far.

Nuclear structure theory is developing rapidly at the moment. Effective field theory is a major framework and one variation of this, halo effective field theory \cite{Ham20}, is not only targeted to describe halo nuclei, but can also now be used to study their beta decays. Care must of course be taken to include the weak interactions in a consistent way, one example being the calculations of the decay of $^6$He into the ground state of $^6$Li \cite{Vai09}.

\section{Experimental procedures}
The field of halo physics has benefitted strongly from the general significant experimental improvements during the last decades in the field of exotic nuclei, both what concerns production methods and detection capabilities. Nuclei close to the proton dripline tend to be easier to produce than nuclei at the neutron dripline, but obtaining sufficient statistics remains a problem for most halo nuclei.

The halflifes of halo nuclei are mostly below 100 ms, $^{11}$Be being the main exception (apart from the deuteron), so production and manipulation of the radioactive nuclei needs to be fast. Isotope Separation On-Line (ISOL) techniques tend to give higher yields than in-flight production for more stable nuclei, but extraction in ISOL systems often happens on timescales of 100 ms or longer which favours in-flight experiments. Decay studies are performed on stopped samples and a thin sample is needed if low-energy emitted particles must be distinguished. The samples can be quite thin for ISOL beams but are typically rather extended for in-flight beams unless a gas catcher step is inserted, which adds extra time delay. Implantation into a detector is therefore employed often with in-flight beams and can, when an ``active target'' is used, retain sensitivity to low-energy particles in much of the stopping volume. A further advantage of implantation studies is that the number of decaying nuclei can be counted in the implantation stage, whereas it has to be established by independent measurement from ISOL/gas catcher samples.

When beta-delayed charged particles are of interest, care must be taken to assure that all relevant events can be measured cleanly. In contrast, detection of gamma-rays or beta-delayed neutrons can proceed efficiently also from extended sources. The beta-particles themselves have an intermediate range in matter and can in some cases constitute a disturbing background for heavier charged particles. One option is then to turn to detectors with reduced beta sensitivy, an example being the optical TPC used in a recent $^6$He measurement \cite{Pfu15}. Detection techniques have now progressed to a point where simultaneous detection of three different particles in the final state is feasible in many cases -- examples include the three final $\alpha$ particles in the decay of $^{12}$N \cite{Dig09} or the $\beta\gamma$n coincidences for the decay of $^{11}$Li \cite{Hir05} -- the main remaining limitation lies in coincident detection of two or more final state neutrons, although first results have appeared for $^{11}$Li \cite{Del19}

In allowed beta-decay the angular correlation between the beta and the subsequently emitted particles will be minute (of recoil order). However, if the decaying nuclei are polarized there may be a significant directional dependence of the beta particle that can be used to obtain spectroscopic information. A clear example of this is given by the measurement of the $^{11}$Li $\beta$n decay \cite{Hir05}.
Furthermore, the parity violation results in a angular distribution between the beta particle and the neutrino. In light nuclei the beta recoil effect is often noticable, see e.g.\ the discussion for the case of $^8$B \cite{Kir11}, and one may therefore observe a broadening of the line shape of the emitted particles that again can be employed spectroscopically.

\section{Decay of neutron halos}
This section presents the current state of knowledge on the decay of neutron halo nuclei, the following section covers other halo nuclei. There is no general consensus on which nuclei to include as having a halo structure, partly because of insufficient experimental information, so the ones highlighted here have an unequivocal halo structure or have sufficient experimental data available to be otherwise of interest. 
For both this and the next section, an overview of the available experimental data can be found at the ENSDF database at the National Nuclear Data Center (accessible on the web at \url{https://nndc.lbl.gov}). There is at the moment of writing (November 2021) insufficient data on the decay of the halo (candidate) nuclei $^{19}$B, $^{19,22}$C, $^{31}$Ne and $^{35}$Na to discuss possible beta decay effects.

\begin{figure}[thb]
\centering
    \includegraphics[width=10.cm,clip]{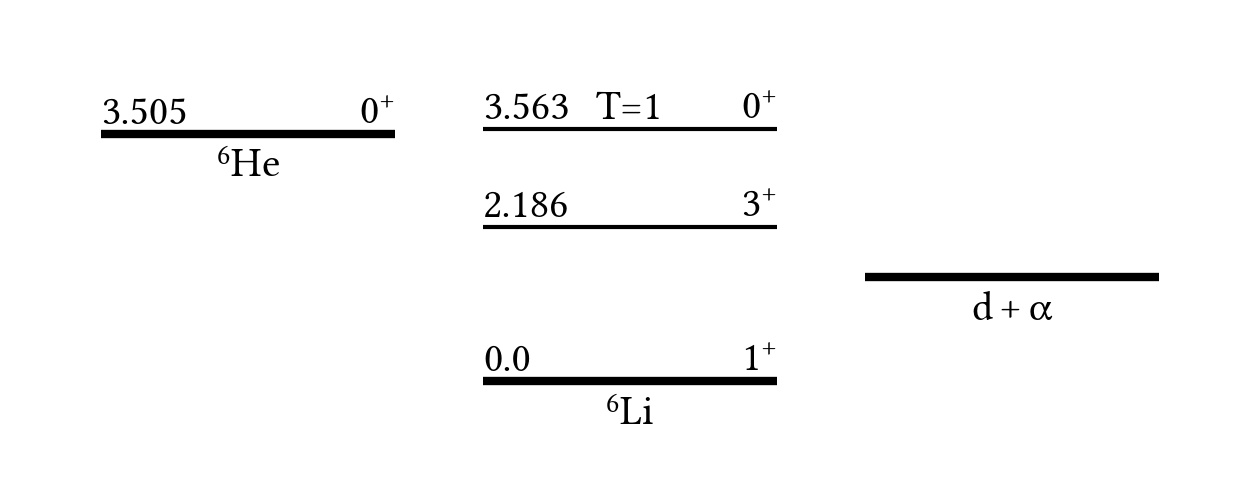} 
    \caption{The beta decay of $^{6}$He leads mainly to the ground state of $^{6}$Li, but a small branch proceeds to the d+$\alpha$ continuum.}
\label{fig:He6scheme} 
\end{figure}

\textbf{$^6$He:}
This is a well-stablished two neutron halo state with the neutrons residing in p-waves, the overall structure was established early on \cite{Zhu93}. The core nucleus is here the $\alpha$ particle that does not beta decay by itself, so the decay of $^6$He is a pure halo decay. Essentially all of the observed beta strength goes to the ground state of the daughter nucleus $^6$Li with a $B(GT)$ of 4.67(2), see \cite{Gne21} for a recent theoretical overview of this transition. The $^6$Li ground state is often treated as a d+$\alpha$ cluster state, it is thus natural to argue that the main halo decay is this ground state transition. However, most efforts have been spent on the beta-delayed deuteron branch that is both experimentally and theoretically challenging, see figure \ref{fig:He6scheme} for the decay scheme.

The first two experiments that observed the $\beta$d branch \cite{Bor93,Rii90} agreed on the spectral shape but differed on its intensity. A high statistics study \cite{Ant02} resolved this and gave a branching ratio of $(1.8 \pm 0.9) \cdot 10^{-6}$ above 350 keV laboratory deuteron energy. The uncertainty on the intensity was reduced considerably by an implantation experiment \cite{Raa09} that gave a branching of $(1.65 \pm 0.10) \cdot 10^{-6}$ above 350 keV deuteron energy. The latest experiment so far employed an optical TPC \cite{Pfu15} thereby achieving a lower detection threshold and giving a branching ratio of $(2.78 \pm 0.07 \pm 0.17) \cdot 10^{-6}$ above 100 keV laboratory deuteron energy. The experiments generally have agreed on the spectral shape that in figure \ref{fig:BGT_betad} has been converted into an experimental $B(GT)$ strength.

The interpretation of the beta-delayed deuteron branch is complex. The ground state can be thought of as a subthreshold resonance (cf. the discussion above on beta-delayed particle emission) and it has a strong coupling to the alpha+d continuum and will therefore in itself give rise to a large strength to the continuum in stark contrast to the experimental result \cite{Rii90}. An ad hoc solution was suggested \cite{Bor93} with interference between (i) beta decay followed by d emission and (ii) 2n emission followed by beta decay, but the situation was quickly clarified by Barker \cite{Bar94} that pointed to the importance of channel decay in R-matrix theory, thereby giving a similar cancellation. Note that this pronounced channel contribution is a clear halo feature.
(Called extra-nuclear contribution in \cite{Rii15}, there also interpreted as leading naturally to the interpretation in terms of decays directly into continuum states.)

With resonance-based theory being this complex it is no wonder that essentially all other calculations of the process have built on the decay-to-continuum approach. There is a corresponding cancellation in these calculations, here between contributions from the inner and outer parts of the wave-functions \cite{Bay94,Tur18}. The two approaches both involve a considerable cancellation giving predictions that are quite sensitive to the input (with ensuing difficulty in making precise predictions) and can in a sense be seen to present similar physics from rather differing platforms.

It is noteworthy that the latest experiment \cite{Pfu15} indeed has a clear indication for an interference effect at low energy, see figure \ref{fig:BGT_betad}, thereby confirming the theoretical interpretations.

\textbf{$^8$He:}
The structure of this nucleus is not completely settled, it is sometimes considered a halo nucleus, sometimes as having a sizable neutron skin. The latest experiment on its decay \cite{Bor93} identify four transitions to excited $1^+$ states in $^8$Li with the highest one at around 9.3 MeV giving beta-delayed tritons and having a $B(GT)$ that according to R-matrix fits \cite{Bar96} is large, in the range 4-5. The neutron branch is also significant, but more work is needed to settle the interpretation of it. Few-body theoretical calculations \cite{Gri96, Shu97} of the decay have also been made and indicated sensitivity to the detailed structure of $^8$He, but there is so far no clear indications for halo signatures in the decay. More experimental data would be welcome.

\begin{figure}[thb]
\centering
    \includegraphics[width=12.cm,clip]{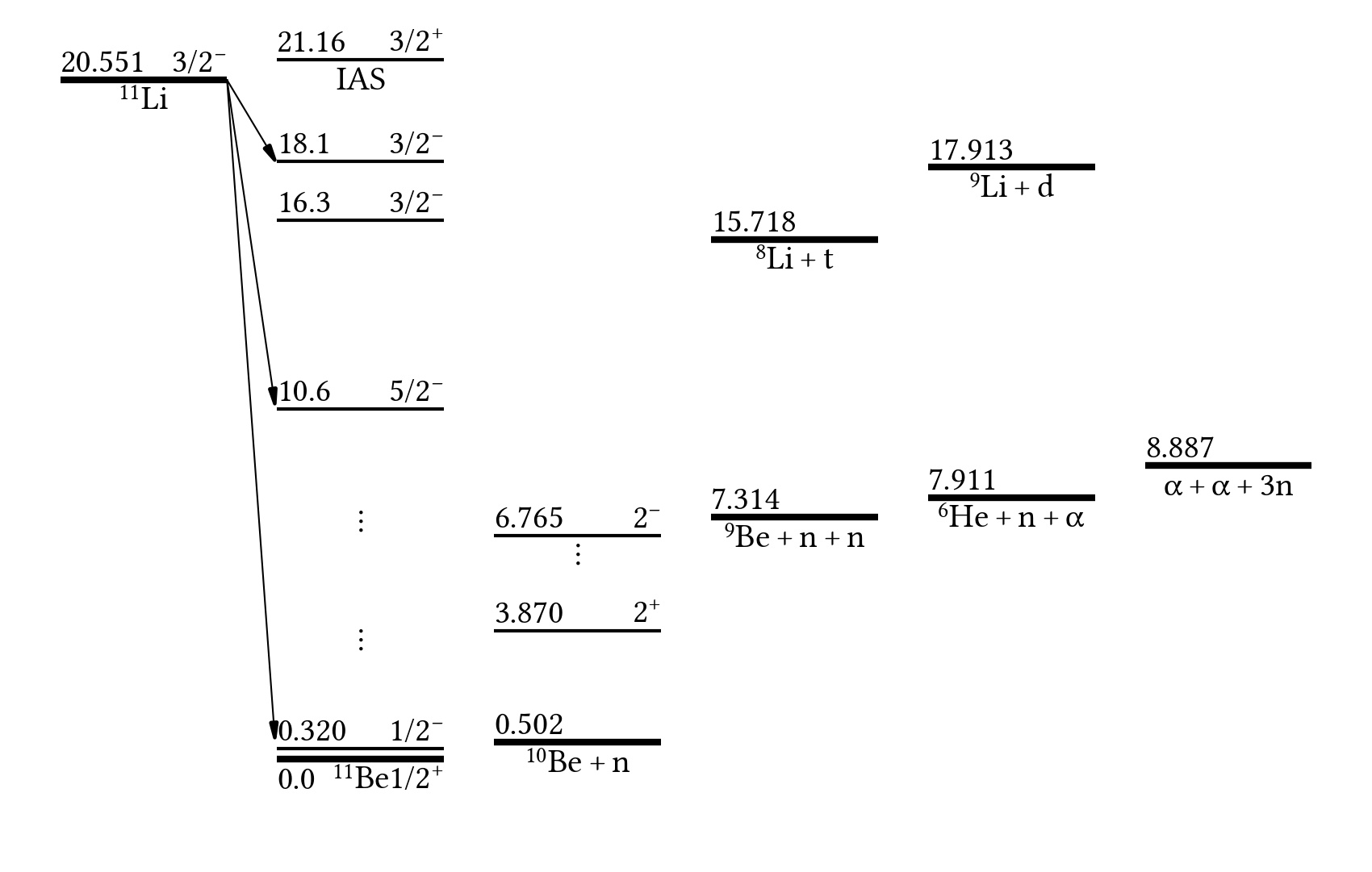} 
    \caption{An overview of the $^{11}$Li beta decay scheme. Note the many identified final channels. More levels in $^{11}$Be and $^{10}$Be are known to play an important part in the decay, but have been left out for clarity.}
\label{fig:Li11scheme} 
\end{figure}

\textbf{$^{11}$Li:}
The beta decay of the important two-neutron halo nucleus $^{11}$Li has been studied in many different experiments. It is a prolific beta-delayed particle emitter with many open decay channels, see figure \ref{fig:Li11scheme}; many decay modes were first observed in its decay. The most extreme branch is the one where the final state after beta decay consists of two alpha particles and three neutrons, a complete break-up. Close to 90\% of the decays go to states up to around 10 MeV excitation energy in $^{11}$Be, but most beta strength lies above this energy range.

Starting from the highest excitation energies, the delayed deuteron branch has now been characterized experimentally \cite{Muk96,Raa08}, see figure \ref{fig:BGT_betad}, with a branching ratio of $1.30(13) \cdot 10^{-4}$ for decays with cm-energy above 200 keV. Several theoretical predictions of the decay have been made \cite{Bay06,Ohb95,Zhu95} and display sensitivity to the final state potential for the deuteron, but the overall agreement with the experimental spectrum is reassuring and there is no doubt that this decay branch is most efficiently described as proceeding directly to the continuum.

Concerning the highest part of the excitation energy range in $^{11}$Be, indications from early studies of the $\alpha\alpha$ and $\alpha^6$He branches \cite{Lan81} as well as the triton branch \cite{Lan84} indicated the presence of a strongly fed resonance just above 18 MeV. A later experiment \cite{Mad08,Mad09,Mad09a} with higher intensity and better detector coverage has confirmed this level at 18.3 MeV and also gave evidence for a second resonance at 16.3 MeV and determined spin-parity for both levels to be $3/2^-$. The next well-established resonance is at 10.6 MeV with spin-parity $5/2^-$ and is also clearly observed in the above channels \cite{Lan81,Mad08}. The triton branch has a branching ratio in the range $(1-2) \cdot 10^{-4}$, the lower limit comes from direct detection in experiments where not all events may have been observed \cite{Lan84,Mad09} and the upper limit from an indirect detection of the daughter nucleus $^8$Li \cite{Muk96}. The total $B(GT)$ strength identified so far in this energy range is around 2.

There is insufficient experimental knowledge on the decays feeding the $\beta$2n and $\beta$3n branches. The important contribution from these branches was established early \cite{Azu80} and the current values of the branching ratios are $P_{2n} = 4.1(4)$\% and $P_{3n} = 1.9(2)$\%. Several possible decay routes have been suggested, but detection of two neutrons simultaneously is rather difficult due to cross-talk between neutron detector units. A recent experiment \cite{Del19} has finally succeeded in obtaing reliable two-neutron coincidences, but the placement of these events in a decay scheme is far from easy. Most neutrons observed in coincidence had energies of a few MeV.

Single neutron branches from levels at lower excitation energy are prominent ($P_{1n} = 86.3(9)$\%) and easier to handle experimentally. The neutrons feed states below the one-neutron threshold in $^{10}$Be, see figure \ref{fig:Li11scheme}, and since the branching to excited states in $^{10}$Be is significant the main information is coming from experiments that also has neutron-gamma coincidence detection \cite{Aoi97,Hir05,Mor97}, the most technically advanced employing polarization of the decaying $^{11}$Li \cite{Hir05}. An alternative to this procedure is to concentrate on precision measurements of the lineshape of the emitted gamma rays \cite{Fyn04,Mat09,Sar04}, an analysis of the recoil broadening of the gamma rays allows deduction of the neutron energies feeding them. The major patterns of the decay to the lowest 9 MeV of excitation energy in $^{11}$Be has been established in this way. Unfortunately, the two most recent decay schemes \cite{Hir05,Mat09} do not agree in detail, so independent measurements would be valuable to clarify the situation. Particular attention has been given to the transitions proceeding through the $2^-$ excited state in $^{10}$Be just 548 keV below the neutron threshold that is a candidate for being an excited s-wave neutron halo state \cite{AlK06}, several of the transitions through the excited $^{10}$Be may be most naturally interpreted as being due to $^{11}$Li core decays \cite{Mat09}.

Finally, the transition to the $1/2^-$ 320 keV bound state has already been mentioned as being useful for determining how much the $^{11}$Li halo neutrons resides in p-orbits \cite{Suz94,Suz97}. Several experiments \cite{Aoi97,Bor97,Mor97} helped to establish the current picture -- consistent with what has emerged from reaction experiments \cite{Sim99} -- that the halo has important s$^2$ as well as p$^2$ components.

Shell model calculations of the $^{11}$Li beta decay has been made by several groups and is also included in many of the above mentioned experimental papers. They agree on placing most of the Gamow-Teller strength at high energy, consistent with the 18 MeV level having large strength, but with the current uncertainties on the experiment decay scheme there has not been detailed comparisons with experiment, nor among the different predictions. From other theoretical frameworks, a quite complete study has been made with antisymmetrized molecular dynamics \cite{Kan10}. There are as yet no other few-body calculations that cover larger parts of the decay scheme.

\begin{figure}[thb]
\centering
    \includegraphics[width=12.cm,clip]{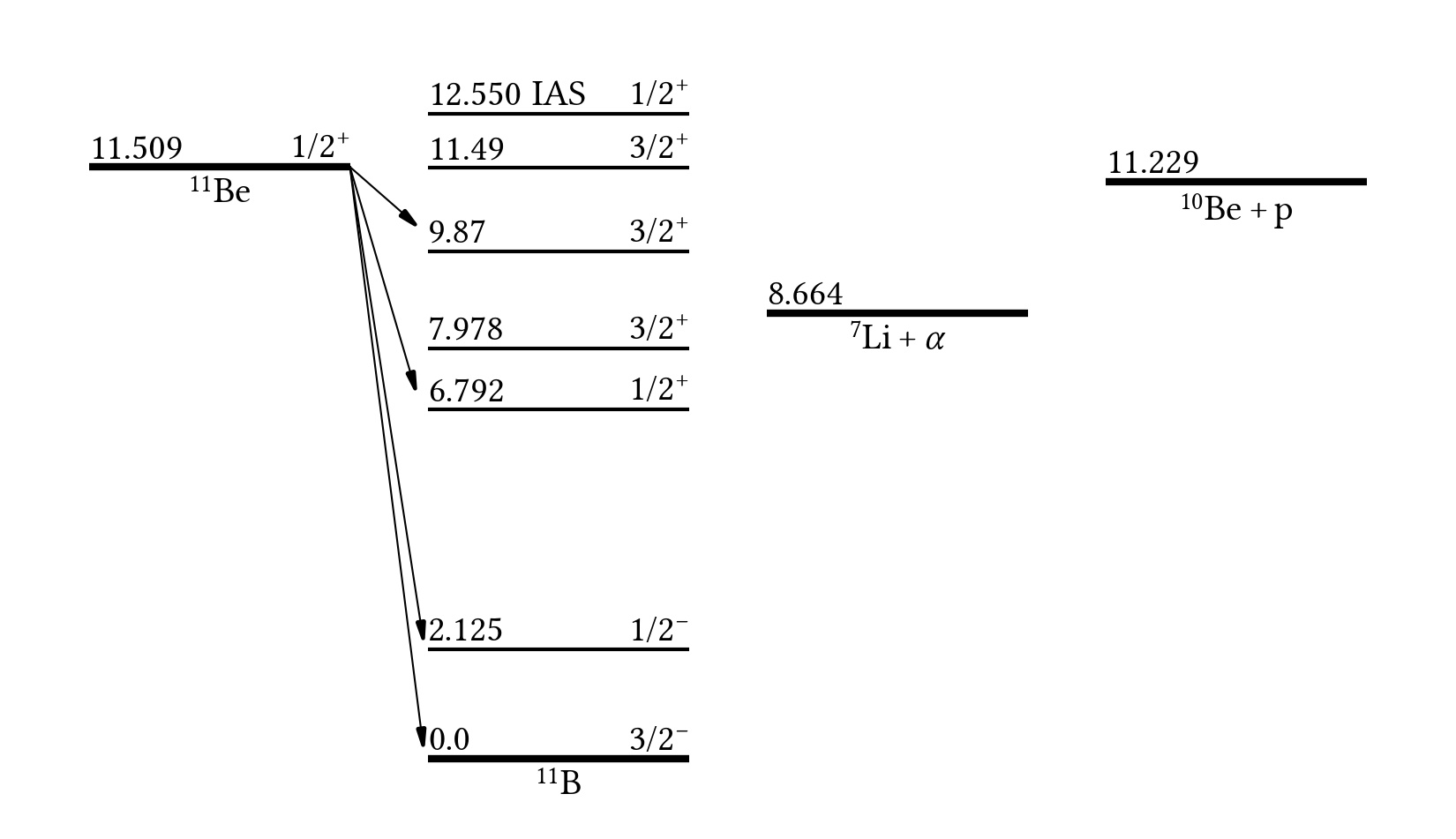} 
    \caption{The main beta decay branches of $^{11}$Be.}
\label{fig:Be11scheme} 
\end{figure}

\textbf{$^{11}$Be:}
Both the $1/2^+$ ground state and the first excited $1/2^-$ state are halos (s$_{1/2}$ and p$_{1/2}$ neutrons outside $^{10}$Be), the ordering being inverted to what the usual filling of orbits would give. This implies that transitions to low-lying states in $^{11}$B will all be first forbidden and gives the overall long halflife of 13.76(7) s (historically this was used to infer that the $^{11}$Be ground state has positive parity \cite{Alb64}). The two transitions with highest beta strength at 9.87 MeV and 11.49 MeV feed states unbound to alpha particle emission, see figure \ref{fig:Be11scheme}. Both states are broad, so the most recent experiment \cite{Ref19} employed an R-matrix analysis to derive a total $B(GT)$ strength of 1.0(2).

The Q-value for beta-delayed proton decay is very small, but observation of this branch would provide a parallel to the beta-delayed deuteron decays of two-neutron halos and several experiments have therefore been carried out \cite{Ayy19,Bor13,Rii14a,Rii20}. Another motivation came from the suggestion that the neutron could decay to a dark particle \cite{For18} in which case one may observe $^{11}$Be decay into $^{10}$Be and a dark (and invisible) particle \cite{Pfu18}. This is relevant if the experiment observes $^{10}$Be rather than the proton, as is the case in all experiments so far except \cite{Ayy19}. A first indication \cite{Rii14a} for the presence of $^{10}$Be could not be confirmed \cite{Rii20} and led instead to an upper limit of $2.2 \cdot 10^{-6}$ of the $\beta$p branch. This is in contrast to the direct observation of a proton line in \cite{Ayy19} corresponding to an excitation energy of 11.425(20) MeV and with an estimated branching ratio of $8.0 \cdot 10^{-6}$. The experimental situation is therefore not settled at the moment.

The theoretical calculations suffer from a lack of knowledge of the potential in the final state between $^{10}$Be and the proton. The first results \cite{Bay11,Bor13} gave branching ratios around $3 \cdot 10^{-8}$, but higher values (up to around $10^{-5}$) may be obtained if a resonance is present in the Q-window for the decay \cite{Rii14,Rii14a}, a conclusion verified by now via several different model calculations \cite{Ayy19,Elk21,Vol20}. A convincing theoretical prediction must reproduce both the $\beta\alpha$ and $\beta$p branches, as attempted with the shell model embedded in the continuum \cite{Oko20,Oko21}. It is clearly of interest to clarify the beta-delayed proton decay experimentally due to the considerable theoretical and conceptual interest.

\textbf{$^{14}$Be:}
The structure of $^{14}$Be is still to be completely clarified and is coupled to the structure of its core (when interpreted as a two-neutron halo) $^{12}$Be. The decay scheme is shown in figure \ref{fig:Be1214scheme}, it is mainly based on experiments published in the period 1995-2002, see ENSDF and \cite{Aoi02,Jep02}.
It seems that essentially all transitions go to a neutron unbound level at 1.28 MeV energy in $^{14}$B (as mentioned earlier, this is strongly resembling the $^{12}$Be decay). Less than 0.6\% of the decay intensity (and probably none) goes to bound states in $^{14}$B, more than 96\% gives one neutron emission mainly through the low-energy $1+$ state with some indication for weak neutron peaks at higher energy, and there is less than around 1\% going to the 2n or 3n branches. Concerning beta-delayed charged particles \cite{Jep02} less than 0.012 \% of the decay intensity gives $\alpha$ particles, whereas tritons were identified corresponding to a branching ratio of 0.02(1) \%. The experiment did not observe beta-delayed deuterons, but a meaningful limit could not be set. Indications for a halo decay component are therefore still absent.

\textbf{Heavier nuclei:}
There are indications for $^{15}$C displaying a moderate one-neutron halo structure, but beta decay is here unlikely to provide much further information. The decay to (particle bound) levels in $^{15}$N is well established.

Among the heavier neutron halo candidates the only nucleus where a reasonably detailed decay scheme has been experimentally established is $^{17}$B \cite{Rai96,Uen13}, but conclusions on possible halo effects in its decay would still be premature.

\section{Decay of other halos}
Proton halos are known in light nuclei, but there are no known states with spatial extension as large as can be found among neutron halos. The effects on their beta decay due to the halo structure are therefore less clear-cut. On the other hand, the hypertriton -- the bound state of a neutron, a proton and a $\Lambda$ particle -- is suspected to have a spatial extension surpassing what is observed in the currently largest neutron halos. It will therefore also be discussed.

\subsection{Proton halos}
The difference in Coulomb energy in mirror nuclei implies that the decay of proton halo states can be compared to that of their mirror nucleus. For the p-wave one-neutron candidate proton halos $^8$B and $^{12}$N that have $T=1$, the mirror decays will even go to the same final states. There are several relevant results for the nuclei with $A=17$, but rather limited ones for even heavier dripline nuclei where the Coulomb effects become stronger so that halo effects are quenched, even for vanishing separation energy.

\textbf{$^8$B:}
This one proton p-wave halo state has received much attention in reaction experiments, but less so in beta decay studies. The main decay branch goes to the broad first excited $2^+$ state in $^8$Be, but most of the beta strength should be concentrated in the 16-20 MeV range that includes $2^+$, $1^+$ and $3^+$ doublets of (partially) isospin mixed states. Only the lowest three of these levels can be reached in weak decays, the two $2^+$ states in $\beta^+$ decay and the lowest $1^+$ state in electron capture. Until recently, only the lowest of these states at 16.6 MeV had been seen in beta decay, but the 16.9 MeV state has now also been observed \cite{Kir11}. The main interest resides in the still unobserved transition to the $1^+$ state at 17.64 MeV; in the terminology used above it could be considered a core decay (the $^7$Be core in $^7$Be+p decaying into the $^7$Li ground state), an interpretation that is consistent with the result of a few-body calculation employing three clusters in the description of each state that predicted rates to all three potentially populated states \cite{Gri00}. The branching ratio to the $1^+$ states is estimated to be around $2.3 \cdot 10^{-8}$ and the first experimental limit on this decay branch \cite{Bor13} is $2.6 \cdot 10^{-5}$ (95\% conf. limit). It will be challenging to reach the needed experimental sensitivity to see the $\beta$p decay.

Reproduction of the experimental alpha particle spectrum from the $^8$B decay within the R-matrix formalism \cite{Bar69,Kir11} (and references therein) has required introduction of at least one extra level beyond the $2^+$ levels at 3 MeV, 16.6 MeV and 16.9 MeV. The nature of the extra level has been discussed repeatedly, see \cite{Rii15} for references to the literature as well as a discussion of the possibility that it reflects decays directly to the continuum.

The IAS will be populated in the decay of proton halo nuclei, but it could be noted that the well established strong isospin mixing between the 16.6 MeV and 16.9 MeV $2^+$ states implies that both will receive Fermi strength. The exact distribution of Fermi and Gamow-Teller strength in the doublet has not been experimentally tested yet.

\textbf{$^{12}$N:}
The last proton here is in a p-wave, has a separation energy of 600 keV and is therefore a halo candidate \cite{Tan13}. Several kinematically complete experiments on its decay have been carried out, the results of the latest analysis \cite{Hyl10} within the R-matrix formalism again indicates a possible contribution from transitions directly to the continuum. Since both this case and the one of $^8$B involves final strongly clustered states (two or three alpha particles), a complete theoretical description of the decay is challenging, but would be illuminating.

\textbf{A=17:}
The interesting beta transition in the mass 17 nuclei is the one from the $1/2^-$ ground state of the two-proton halo candidate $^{17}$Ne to the first excited $1/2^+$ state in $^{17}$F with a structure of an s-wave proton outside $^{16}$O and a separation energy of 105 keV, only. The branching ratio of this first-forbidden transition was measured \cite{Bor93a} to be 1.65(16)\% (confirmed in an independent measurement \cite{Oza98} to be 1.56(20)\%) whereas the strength of the mirror transition would have resulted in a branching ratio of 0.75(12)\%. The factor two difference in strength in the two mirror transitions must be related to the different separation energies for the entering nuclei, and it should be noted that the $^{17}$Ne transition has higher strength than that of the mirror $^{17}$N transition. Different theoretical calculations \cite{Bor93a,Mic02,Mil97} have explored the effects of different wavefunction radial extensions and different occupation probabilities of the s-orbit, it seems that first-forbidden transitions are quite sensitive to model assumptions and more work would be needed to obtain a unique interpretation.

Similar large differences in the strength of mirror beta transitions have been reported for decays of the heavier proton dripline nuclei $^{22}$Si and $^{26}$P \cite{Lee20,Per16}.

\subsection{Lambda halos}
Due to significant technical developments hypernuclei are now starting to become accessible for more detailed experimental studies, a recent overview is provided in \cite{Fel15}. There are in particular several studies from RHIC and LHC where the hypertriton (among many other nuclei) are produced in high-energy heavy-ion collisions.

The hypertriton is clearly a halo, and a two-body halo of a $\Lambda$ particle and a deuteron \cite{Jen04}. However, the binding energy of 0.4(2) MeV \cite{Ada20} is not known sufficiently well to give accurate descriptions of its properties although it is clear that it has a quite extended structure. Experiments on it are hindered by its short lifetime, so very little is experimentally known except for its weak decay that is currently the main source of information.
The main decay branch of the $\Lambda$ is $\Lambda \rightarrow p + \pi^-$.
Corresponding to this, the dominating decay modes of the hypertriton are the 2-body decay $^3_{\Lambda}H \rightarrow ^3He + \pi^-$ and the 3-body decay $^3_{\Lambda}H \rightarrow d + p + \pi^-$.

The hypertriton lifetime, which is used rather than the halflife in particle physics, was due to the spatial extension expected to be close to that of the $\Lambda$ particle, 263(2) ps \cite{PDG20}. During the last decade the lifetime has been measured with increasing accuracy, leading to values ranging from below 150 ps to around 240 ps but mostly staying well below the $\Lambda$ lifetime. The current average for the hypertriton is 200(13) ps \cite{Abd21} (the paper gives an overview of the experimental and theoretical situation and offers an explanation in terms of final-state interactions). It will interesting to follow the steady improvements in hypernuclear experiments and see what other halo-relevant information can be obtained.

\section{Summary and outlook}
Beta decay is a valuable probe of nuclear structure in general, partly due to the simple structure of the operators, eq. (\ref{eq:operators}), for allowed transitions, partly due to the clean selection rules that follow from these operators. It has proven to be a versatile probe also of the structure of halo nuclei, the main challenge being that the beta decay phase space factors favour transitions to low-lying states in the daughter nucleus, whereas much of the interesting physics resides in transitions with relatively large values of $B(GT)$ that tend to occur at higher excitation energies in the final state.

The clearest observed signature for the halo structure is the beta-delayed deuteron emission that is now established for $^6$He and $^{11}$Li (one should not forget that the main $^6$He decay into the ground state of $^6$Li may also be interpreted as a two-neutron to deuteron transition). It may be expected that other decay branches will also give signatures for beta transitions proceeding directly to continuum states, but obtaining a unique signature for this decay mechanism may take time. The conceptual division into decays of the halo particles and the core is in some cases a useful tool to describe the overall decay patterns, but this division cannot be taken too far since isospin appears to be still very well conserved in halo nuclei. Finally, beta decay has in several cases produced very useful constraints on the detailed configuration of halo states.

The neutron halos at the neutron dripline will all, apart from $^6$He and $^{11}$Be, have large branching ratios for beta-delayed one neutron emission as well as a possibility for beta-delayed emission of two or more neutrons. The latter remains an experimental challenge, but it is noteworthy that there is so far very little theoretical work on the multi-neutron branches to guide the interpretation of the limited data.

In studies of the decay of proton halos it is more straightforward to measure all emerging radiation (except of course the neutrinos). However, so far fewer clear halo features have been observed. For the light p-wave proton halo candidates $^8$B and $^{12}$N a strong alpha-particle clustering appears in the populated final states, giving broad structures in the final states that makes it more difficult to interpret the spectra without using more complex frameworks such as the R-matrix. It is important to have such well-studied cases, since there are first indications for similar types of structure to appear in the decay of $^{11}$Li. A more extensive theoretical modelling of the decays leading to multi-particle final states would be very welcome.

The decay scheme is only well established for a few of the known halo nuclei, mainly proton halos, $^6$He and (apart from the $\beta$p branch) $^{11}$Be. It will be interesting to follow how the continuing experimental development will test the current understanding of how the halo structure influences the beta decay process.

\end{document}